\documentclass[a4paper,openany, 12pt]{book}

\usepackage[text={15.5cm,23cm},centering]{geometry}
\usepackage[sectionbib]{chapterbib} 
\usepackage{mathptmx}
\usepackage{graphicx}
\usepackage{subfigure}
\usepackage{color}

\newcommand{\chapterauthor}[1]{\textsc{#1}\section*{}}  

\usepackage{outlines} 
\usepackage{soul}
\usepackage[dvipsnames]{xcolor}
\usepackage[round]{natbib}
\usepackage[hyperfootnotes=false]{hyperref}
\hypersetup{
  colorlinks,
  citecolor=Violet,
  linkcolor=Violet,
  urlcolor=Blue}

\usepackage{amssymb}

\newcommand\new[1]{\textcolor{Black}{#1}}
\newcommand\final[1]{\textcolor{Black}{#1}}

\newcommand\mandt{\citealt{Mandt2025}}
\newcommand\bauer{\citealt{Bauer2025}}
\newcommand\sickafoose{\citealt{Sickafoose2025}}
\newcommand\woodney{\citealt{Woodney2025}}
\newcommand\kareta{\citealt{Kareta2025}}

\newcommand\fv{\citealt{FV2025}}

\begin{document}

\setcounter{chapter}{6}

\chapter{Evolutionary Processes in the Centaur Region}

\chapterauthor{Rosita Kokotanekova, Aurélie Guilbert-Lepoutre, Matthew M. Knight, Jean-Baptiste Vincent}



Centaurs populate relatively short-lived and rapidly evolving orbits in the giant-planet region and are believed to be one of the solar system's most complex and diverse populations. Most Centaurs are linked to origins in the dynamically excited component of the trans-Neptunian region, and are often considered an intermediate phase in the evolution of Jupiter-family comets (JFCs). Additionally, the Centaur region hosts objects from varied source populations and having different dynamical histories. In this chapter, we focus on the physical processes responsible for the evolution of this heterogeneous population in the giant-planet region. The chapter begins with a brief review on the origin and early evolution that determine Centaurs' properties prior to entering the giant-planet region. Next, we discuss the thermal, collisional, and tidal processes believed to drive the changes Centaurs undergo. We provide a comprehensive review of the evidence for evolutionary changes derived from studies of the activity, physical properties, and surface characteristics of Centaurs and related populations, such as trans-Neptunian objects, JFCs, and Trojans. This chapter reveals a multitude of gaps in the current understanding of the evolution mechanisms acting in the giant-planet region. In light of these open questions, we conclude with an outlook on future telescope and spacecraft observations, detailing how they are expected to elucidate Centaur evolution processes.

\section{Introduction}

\label{sec:intro}
Centaurs are a population of small solar-system bodies which orbit the \new{S}un in the realm of the giant planets,
\new{but} outside of 1:1 mean motion resonances with \new{them.}
These objects populate chaotic orbits with short dynamical lifetimes estimated from a few Myr up to tens of Myr (see \citealt{DiSisto2025}{ for a review}). 
\new{As was introduced in \cite{Volk2025}{}}, the existence of the population requires the Centaur region to be continuously replenished with objects from other reservoirs. The main source of Centaurs has been identified as 
\new{the dynamically excited trans-Neptunian population \citep[\final{consisting of} the dynamically excited or hot classical Kuiper Belt objects (KBOs), the Scattered Disk, and various resonant populations below 50~au, e.g.,][]{nesvorny2017}}. 
At the other end of Centaurs' dynamical evolution, objects which do not get ejected typically evolve towards the inner solar system as Jupiter Family Comets (JFCs). This characteristic pathway defines Centaurs as an intermediate evolutionary phase between the Trans-Neptunian objects (TNOs) and comets.

Since the discovery of the Kuiper Belt in the 1990s, objects from the TNO-Centaur-JFC continuum have been regarded as valuable probes preserving evidence for the conditions in the early solar system. This promising lead was among the main arguments supporting the spectacular lineup of space missions to comets \citep{Snodgrass2023}\new{, NASA's New Horizons exploration of the Kuiper Belt \citep{Stern2015,STern2019}, as well as the recently proposed Centaur mission concepts described in \cite{Harris2025}}.
Space missions 
along with ground-based observational and theoretical advances of the past decades have resulted in a new, more complex view on these populations. It is now widely accepted that JFCs and Centaurs have undergone significant processing throughout their history since formation \citep[\new{e.g.,}][]{gkotsinas2022,Pajola2023,Filacchione2022} \final{and should rather be perceived as complex geological bodies.}. Naturally, this poses challenges to the notion that we can derive unambiguous constraints on models of the early solar system. On the other hand, 
it opens the possibility to acquire \new{a} better understanding of the rich variety of evolutionary processes \new{thought} to have shaped today's populations.

Centaurs are key for making progress on this task, but to date they remain the least understood population of the TNO-Centaur-JFC continuum. 
\new{Moreover, the Centaur population also provides a valuable resource for understanding other small-body populations. Centaurs cannot be viewed simply as a phase of JFC evolution. In addition to perturbed TNOs, the Centaur region is \new{thought} to host objects originating from the Oort Cloud, as well as the Trojans, Hildas and the outer main asteroid belt (see Section \ref{sec:life-cycle-dynamics} and \citealt{DiSisto2025}). The reader should therefore keep in mind throughout the chapter that the Centaur region can be regarded as an agglomeration of objects with potentially different formation histories and different levels of evolutionary processing.}

In this chapter we review the current understanding of the processes governing the evolution experienced by these diverse objects in the giant planet region. The processes described below have been observed or proposed to act on Centaurs
to produce physical and chemical changes to the surface layers of individual objects, and as a result \final{to} shape the cumulative population properties. In order to focus on discussing processes characteristic \new{of} the giant planet region, we will generally stick to the strict Centaur definition requiring orbits enclosed in the giant planet region (q$\geq$ $a_\mathrm{Jupiter}$, Q$\leq$ $a_\mathrm{Neptune}$, see \new{\citealt{Volk2025}{}}).


The current review of Centaur evolution is a natural continuation of an increasing body of work from the past few decades. However, past works predominantly considered Centaurs in the context of their source populations in the Trans-Neptunian region (see \citealt{Davies2004} and \citealt{Barucci2008}) while active Centaurs, in particular, were reviewed together with comets (see \citealt{Festou2004}). Notably, an increasing number of works have focused on discussing Centaurs as part of the TNOs-Centaurs-JFCs continuum (see \citealt{levisonduncan1997,Jewitt2015,Peixinho2020,Jewitt2022,Fraser2023}).
This volume is the inaugural standalone review of Centaurs and this chapter offers the first opportunity to spotlight Centaurs' evolution processes.

To this end, the chapter aims to combine evidence from all aspects of Centaur research in order to review our current understanding of the processes evolving Centaur surfaces and upper layers. This paper is therefore closely related to several other chapters in this issue. In particular, we would like to draw the reader's attention to the chapters which provide comprehensive reviews on the topics that are key for understanding Centaur evolution. We refer the reader to {\cite{Johansen2025}} and {\cite{DiSisto2025}} for comprehensive reviews on the formation and dynamical evolution of Centaurs. Reviews of the observational evidence discussed in this paper can be found in: {\cite{Fernandez2025}}, focusing on the physical properties of Centaurs; \cite{Peixinho2025}, surface composition; {\sickafoose}, nearby environment; and {\cite{Mandt2025}}, volatile composition. We would like to highlight that this chapter is closely related to \cite{Hirabayashi2025}. While \cite{Hirabayashi2025} focus on inferring the interior properties of Centaurs, the current chapter explores the contribution that evolutionary processes in the giant planet region have on Centaur surface layers and on the overall processing/aging of the population.

The review is structured as follows: in Section~\ref{sec:life-cycle} we provide an overview of the dynamical pathways of Centaurs and their links to other solar-system populations. Section \ref{sec:processes} reviews the physical processes which are \new{thought} to drive the evolution of Centaurs in the giant planet region, followed by Section \ref{sec:evidence} where we present the observational evidence motivating and constraining the physical models of Centaur evolution. We conclude the chapter by discussing the advances of Centaur evolution we foresee for the coming years (Section \ref{sec:outlook}) and a chapter conclusion (Section \ref{sec:conclusion}).

%
%

\section{The Life Cycle of Centaurs}
\label{sec:life-cycle}

In this chapter, we explore the evolutionary processes that shape the Centaurs in the present epoch. This, however, is a notoriously complex endeavor. The Centaur region is a melting pot of small bodies with different origins, dynamical histories, and past processing. Each object reflects not only the conditions at its formation in the early solar system, but also the subsequent alterations affecting it throughout its lifetime. We begin our effort to disentangle the complexity of this region with a short census of the Centaur region. We discuss the life cycle of Centaurs and identify the other small-body populations likely to provide key evidence for Centaur evolution. 
In the interest of space, this review omits a number of details\final{,} \new{such as specific formation mechanisms and the exact pathways of dynamical evolution\final{,}}
which the interested reader can find in other chapters (see \citealt{Johansen2025}{} and \citealt{DiSisto2025}).

\subsection{\new{Origins} of Centaurs} 
\label{sec:life-cycle-dynamics}

\begin{figure}[h]
\label{fig:life-cycle-diagram}
\begin{center}
\includegraphics[width=0.99\textwidth]{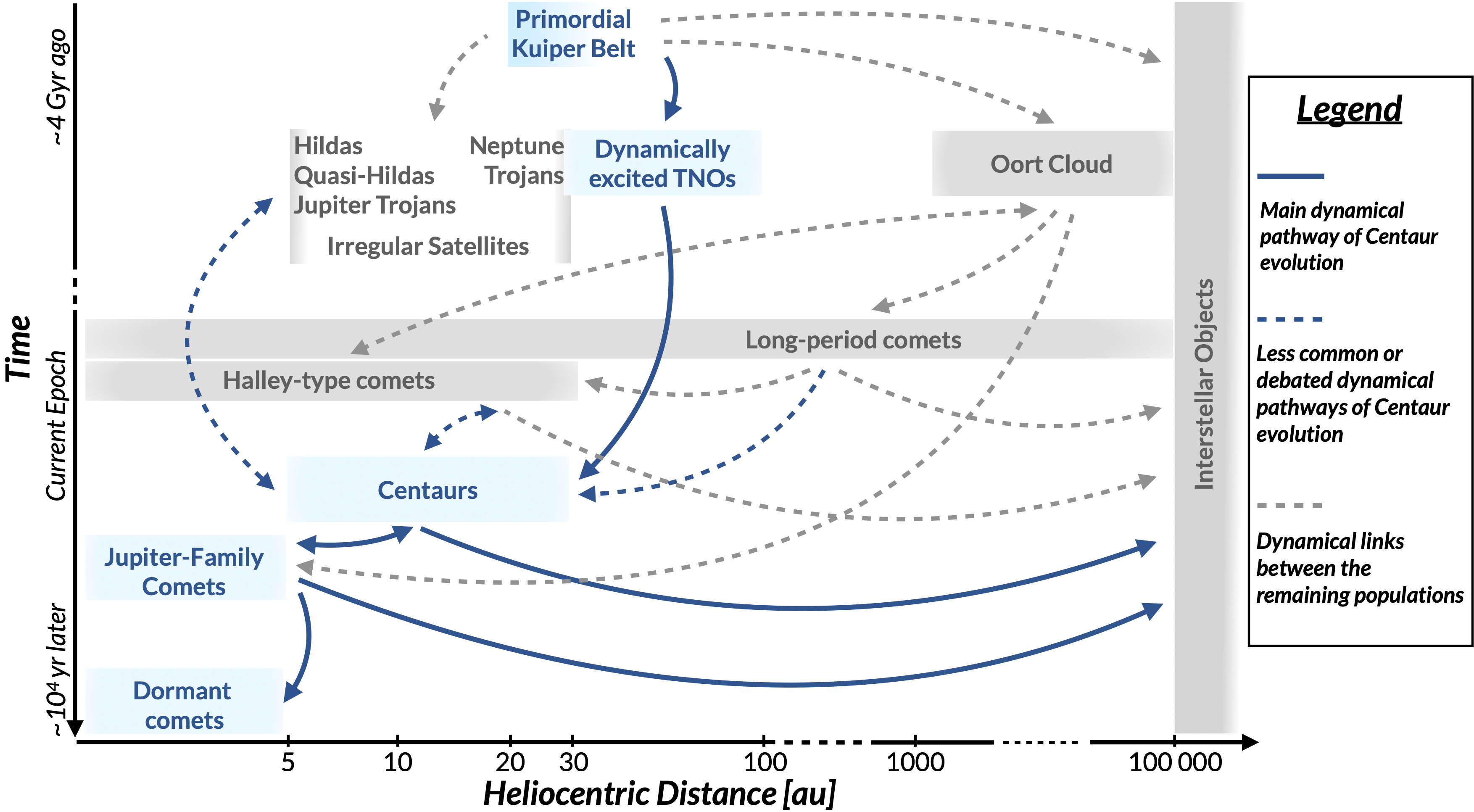}
\end{center}
\caption{Schematic representation of the dynamical evolution of Centaurs and the minor-planet populations related to them. 
\new{This diagram highlights the main pathway of the TNOs-Centaurs-JFCs continuum (blue arrows) and depicts the complex dynamical links of Centaurs with other small-body populations. \final{The approximate range of typical orbits is shown by the horizontal extent of a given box.} The blue boxes indicate the populations representing the main stages of the Centaur life cycle. In the cases when the population density and/or the definition of a given population varies across different works the shading of the boxes indicates the uncertainty in the heliocentric range. For simplicity, Hildas, Quasi-Hildas, Trojans and irregular satellites have all been depicted together within the heliocentric distance range of the giant planet orbits.}}
\end{figure}

The provenance of Centaurs can be traced back to a formation in the outer planetesimal disk at $\sim$20-30 au from the Sun \cite[see][and references therein]{Nesvorny2018}. Following the gas disk dissipation (within the first 10 Myr of the solar system formation), this region, also known as the primordial Kuiper Belt (PKB), is \new{thought} to have undergone
extensive dispersal. Our understanding of this period $\sim$ 4.2 Gyr ago comes predominantly from the family of planetary instability models  \cite[see][for a review]{Nesvorny2018} derived from the original Nice model by \cite{Tsiganis2005}. 
According to these dynamical simulations,  most objects from the PKB \new{(with the exception of Cold Classical TNOs which have remained on approximately the same orbits since formation)} were ejected into interstellar space. \new{A small fraction of the perturbed planetesimals remained} in the solar system and populated today's asteroid belt, Jupiter and Neptune Trojans, irregular satellites of the giant planets, dynamically excited TNOs, and the Oort cloud (Fig. \ref{fig:life-cycle-diagram}) 

At the current epoch, a variety of destabilizing mechanisms can act on objects stored in the Trans-Neptunian region \citep{nesvorny2017} to force them back into the inner regions of the solar system.
Due to gravitational interactions with the giant planets, they enter a regime of chaotic motion which gradually reduces their perihelion distance until they reach low-inclination Neptune-crossing orbits, followed by a dynamical cascade supplying the giant planet region with icy objects now called Centaurs \citep{duncanlevison1997}.
Pathways from the outer to the inner solar system can be very complex \citep[e.g.,][]{Fraser2023}, described by two modes of migration through the giant planet region: either dynamical chaos or mean motion resonance hopping \citep{baileymalhotra2009,seligman2021}. 
These pathways can be fast or rather lengthy if the object remains trapped in the Uranus-Neptune region, with characteristic dynamical timescales reaching up to \new{gigayears} 
\citep{disistobrunini2007}. On average however, Centaurs remain in the giant planet region for \new{$\sim$}10~Myr \citep{levisonduncan1997,tiscarenomalhotra2003,volkmalhotra2008,baileymalhotra2009}.
Each Centaur thus follows a unique and distinct dynamical track to reach the orbit where it is currently observed.
When the orbit becomes controlled by interactions with Jupiter, multiple passages from the giant planet region to the inner solar system (where Centaurs would typically be called Jupiter-Family Comets) are possible \citep{sarid2019,seligman2021,guilbert2023gateway}.
This constitutes the main dynamical pathway of Centaurs and gives rise to the most prevalent  view of Centaurs as part of the ``TNO-Centaur-JFC continuum.'' Even though the TNO region is considered the main Centaur reservoir \citep{Disisto2020}, the Centaur region is likely to also host objects sourced from other populations. Minor contributions are \new{thought} to originate from Jupiter Trojans \citep[see][and references therein]{Disisto2019}, Neptune Trojans \new{\citep[e.g.,][]{Horner2010}}, 
Quasi-Hildas \citep{GilHutton2016}, and Haumea family members \citep{Lykawka2012}. Other possible source populations, specifically of high-inclination and retrograde Centaurs \citep[see][]{Disisto2020, Kaib2023} are proposed to be Oort-Cloud comets \citep{Brasser2012,Fouchard2014,delaFuenteMarcos2014} and  escaped asteroids from the inner solar system \citep{Greenstreet2020}.

Most, if not all, of these Centaur source populations can be traced back to a formation in the PKB \citep[see][]{Nesvorny2018}. It is, therefore, plausible to expect that all Centaurs still share some common properties. 
\new{However, the different Centaur source populations have evolved in different parts of the solar system and have therefore undergone various levels of collisional and/or thermal processing before entering the giant planet region. Once in the Centaur region, the heterogeneity is expected to increase further due to the differences in the individual objects' dynamical paths. This suggests heterogeneity among Centaurs which goes beyond the established classifications such as active/inactive (see \bauer{}) and red/neutral Centaurs (see Section \ref{sec:evidence-colors}) To understand this complexity, it is therefore important to consider the processes which have determined the properties of incoming Centaurs. }

\subsection{Early evolution}
\label{sec:early-evolution}

\new{In this section, we briefly review the early evolution of Centaurs' progenitors within the PKB, during the epoch of planetary migration and subsequently within the Centaur reservoirs.} The formation and very early evolution of PKB planetesimals \new{prior to and during planet migration} determined the initial chemical, physical, and thermal properties of the Centaur parent populations. 
Constraining the degree of processing sustained during the earliest stages of a Centaur’s evolution is extremely difficult. To provide a quantitative assessment, we would need to have identified their formation mechanism, the time at which these objects formed, and their initial composition. Most of our current understanding thus relies on theoretical expectations derived from models which are \final{based on the properties of the currently observed objects and populations,}
and are applied to explore the free parameter space corresponding to the PKB and the outer solar system  reservoirs. 
\new{Bearing these caveats in mind, in the following,} we outline a few aspects of the early evolution of Centaur progenitors which are most relevant to the discussion of today's evolution in the giant planet region.

\subsubsection{Early thermal evolution}

Thermal processing during the early solar-system stages includes internal heating by radioactive elements \citep[e.g.,][]{Prialnik1987,Haruyama1993}. The possible outcomes of this process strongly depend on the size of \new{the objects},
their thermal conductivity, their composition (in particular the dust-to-ice mass ratio), and the timing of their formation. With low thermal conductivities, heating is more effective and deep interiors are thus more likely to reach temperatures triggering phase transitions \citep[e.g.,][]{Prialnik1995}. 
Additional sources of energy can also be considered: for example accretional energy,
\new{although most likely insignificant for most small objects, could yield more diversity to the expected outcomes if added to radiogenic heating \citep[e.g.,][]{Merk2006}}.
The aforementioned studies suggest that it is more probable to find pristine material in the outermost layers of icy objects rather than in their deep interiors. 
By considering the possibility of collisional heating in addition to radiogenic heating, \citet{Golabek2021} estimated that the preservation of highly and moderately volatile species such as CO and CO$_2$ would only be possible for objects smaller than 20~km that formed relatively late ($>$3.5~Myr).

Finally, insolation remains a prevalent cause for thermal processing. Recently, simulations by \citet{Davidsson2021a}\new{, \cite{Steckloff2021}, and \cite{Parhi2023}}
suggested that all icy objects would lose their highly volatile content present in the form of pure ice on timescales ranging from a few thousand to 200~Myr.
Those species would need to be trapped in a more stable phase (like water ice or CO$_2$ ice \new{in amorphous or crystalline form}) to survive on longer timescales. 
We expect that chemical differentiation should take place \new{on longer timescales}, as equilibrium temperatures met in the trans-Neptunian region \new{($\sim$30-50~K)}
exceed the sublimation temperature of pure icy species like CO, N$_2$, or O$_2$ \citep{DeSanctis2001,Choi2002,Merk2003,Malamud2015,Kral2021,Lisse2021,Loveless2022,Malamud2022,Parhi2023}.
As a consequence, Centaurs which originate from the Kuiper Belt and the Scattered Disk are expected to be depleted to some degree of their most volatile free-condensed \new{(e.g., pure ice, not trapped)} species,
within at least a subsurface layer of a few kilometers, 
even before they enter a pathway that brings them toward the inner solar system. The fact that such species are currently observed in active Centaurs (see \fv{} and \mandt{}) indicates that we do not yet have a full understanding \new{of how the most volatile species are retained for the age of the solar system}.

\subsubsection{Early collisional evolution}

Another highly-debated aspect of the early evolution of Centaur progenitors is their past collisional evolution. On one hand, the characteristic physical properties of JFCs (e.g., their shapes, high porosities, low tensile strength, high volatile content, etc.) have been interpreted as clear signatures of their \new{primitive} nature \cite[e.g.,][]{Davidsson2016} 
\new{On the other hand,}
there is building evidence that JFCs (and therefore Centaurs as their progenitors) are the outcome of collisional processing in the PKB, during the disk instability phase and while residing in the Scattered \new{D}isk \citep{Morbidelli2015,Bottke2023}. Furthermore, in support of this view, it has been shown that 
\new{properties that were previously thought to exclude comets from being collisionally evolved, such as high porosities and the presence of volatiles,}
can survive catastrophic disruptions \citep[e.g.,][]{Jutzi2017,Schwartz2018,Steckloff2023} 
For more details on the subject of early collisional evolution, we refer the reader to \cite{Fernandez2025}.

\subsubsection{Space weathering}
\label{sec:space_weathering}

During their residence in the outer solar system, icy objects are subject to long-term irradiation by solar wind, 
\new{UV photons, and cosmic rays, as well as by the impacts of micrometeorites.}
The effects of such continuous bombardment 
\new{are known to affect the surface composition, color, albedo, and structure of the}
surface 
\new{material}
\citep[e.g.,][]{Hapke2001,Cooper2003,Hudson2008,Pieters2016}.
\new{Prolonged ion irradiation can also act to amorphize water ice \citep[e.g.,][]{Dartois2015}.}
The impact \new{of space weathering} on surface chemistry strongly depends \new{on} the initial 
composition, since the net effect is to break molecular bonds to produce irradiation products in the form of new surface and subsurface molecules.

\new{Centaurs are subjected to space weathering during their prolonged residences in their reservoirs. Therefore, it could be expected that the surfaces of recently escaped Centaurs} 
should reflect the influence of long-term irradiation\new{,} 
unless they are modified by some \new{other} process 
\new{once they enter the} giant planet region (see Section \ref{sec:processes}). \new{However, i}t is not \new{considered likely}
that space weathering \new{during} the Centaur phase is responsible for noticeable surface evolution. Due to its location in the middle heliosphere, the Centaur region experiences moderate irradiation from solar energetic ions (\new{whose flux} 
decreases with increasing heliocentric distance) 
and galactic cosmic rays and ion fluxes from the interstellar environment (which 
\final{increase}
further away from the Sun). For the radiation dose at $\sim$1 $\mu$m depth for objects at 20~au estimated by \cite{Hudson2008}, and the saturation doses for volatile ice irradiation found by \citet{Brunetto2006}, the irradiation timescale (i.e., the time needed to change the properties of the material down to a given depth) is between 50~Myr and 1~Gyr \cite[see][]{Wong2016}. Similarly long timescales of $10^8$ years have been estimated for forming a meter-deep irradiation crust depleted of volatiles \citep{ShulMan1972,Luu1996}. These time scales are longer than the typical dynamical lifetime in the Centaur region ranging from a few Myr up to tens of Myr \cite[][and references therein]{Peixinho2020} and the really short blanketing timescales 
\new{(the time over which a comet's surface is covered by a layer of fallback coma material, i.e.,\citealt{Jewitt2002,Jewitt2015}).}
It is therefore unlikely that space weathering drives the surface evolution during the Centaur phase and it is not discussed further in \final{the following section.}



%
%

\section{Processes driving the evolution of Centaurs}
\label{sec:processes}

\subsection{Thermal Processing}
\label{sec:proc-thermal}

Thermal processing is identified as the main driver of Centaur evolution in the giant planet region. The course of an object's thermal evolution is tightly connected to the long-term orbital evolution\new{,} which
determines its thermal environment. 
As insolation increases with decreasing heliocentric distance, Centaurs periodically reach regions where surface and 
\new{subsurface temperature may favor various phase transitions (described below) that may result }
in observable cometary activity. \final{As of now,} 
Centaurs with perihelia located beyond Saturn appear to lack detectable activity \citep{cabral2019,li2020,lilly2021}, which is generally taken as a sign that sublimation of 
\new{hypervolatile}
species such as CO or CH$_4$ 
is not driving the activity of Centaurs at these distances \final{in contrast to what has been inferred for Long period comets \citep[e.g.][]{Jewitt2021}}.
Crystallization of amorphous water ice has instead been 
\new{proposed}
as an efficient source of activity 
\citep{Jewitt2009,prialnikjewitt-comets3}, at least in the 5-10~au region and possibly up to 12~au \citep{guilbert2012}. 
\new{Beyond about 10 au, crystallization rates strongly decline and amorphous water ice crystallization is thought to be less efficient than sublimation of CO$_2$  ice or the sublimation of segregated hypervolatile species from CO$_2$ ice \citep{davidsson2021}. 
The latter process, in which hypervolatiles such as CH$_4$ \citep{luna2008}, N$_2$ \citep{satorre2009}, or CO \citep{simon2019} are trapped in CO$_2$ ice and released at temperatures above their individual sublimation temperatures, is similar to the trapping and release of molecules during the crystallization of amorphous water ice.}

\begin{figure}[t]
\begin{center}
\includegraphics[width=\textwidth]{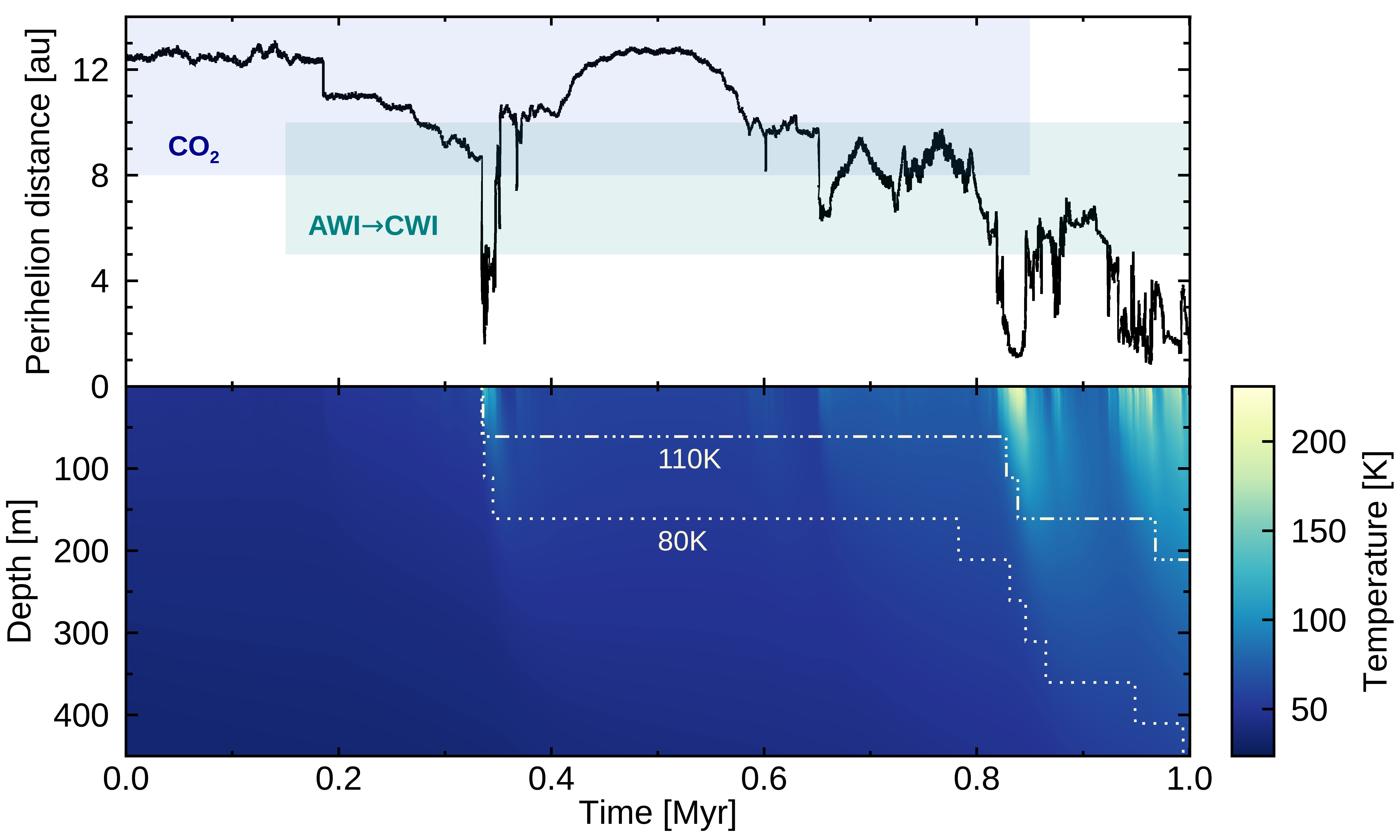}
\end{center}
\caption{Example dynamical and resulting thermal evolution for an object leaving the trans-Neptunian region, reaching the giant planet region, and eventually reaching a JFC orbit \citep{nesvorny2017, gkotsinas2022}. The top panel shows the evolution of the perihelion distance as a function of time for the last Myr of evolution before the object is ejected out of the solar system. \new{Shaded areas illustrate the most likely drivers for Centaurs’ activity: CO$_2$ segregation or sublimation (in blue), amorphous water ice (AWI) to crystalline water ice (CWI) transition (in green). We note that these phase transitions can also occur at lower heliocentric distances.} The bottom panel shows the distribution of the internal temperature as a function of depth and time, resulting from the orbital evolution. The depth\new{s} of two isotherms \new{are} given as guides: dotted line for 80~K and dash-dotted line for 110~K. 
}
\label{fig:orbit-evolution}
\end{figure}

As the heat wave and phase-transition fronts progressively move inward below the surface of Centaurs (see an example in Fig.~\ref{fig:orbit-evolution}), the composition of the top layer could change to a degree that prevents any further activity. For example \citet{guilbert2012} suggested that activity driven by the crystallization of amorphous water ice could only be sustained for 10$^4$~yrs at most, a small fraction of the typical dynamical timescale. 
\new{This implies that Centaurs that are currently active}
should have recently undergone a sunward orbital change 
\new{\citep[e.g.,][]{Lilly2024}}
allowing for phase transitions to be triggered deeper below the surface and producing some observable cometary activity. Empirical data seem to confirm that active Centaurs \final{have undergone drastic drops in their perihelion distances within the past 10$^2$ to 10$^3$~yrs \citep{Fernandez2018,Lilly2024}.} 

The emerging picture is that the activity of Centaurs should be closely related to their orbital evolution. For example, \citet{cabral2019} argued that crystallization-driven activity should be observed for objects dynamically new to the Jupiter-Saturn region, as others having previously stayed in this region would have exhausted their amorphous water ice content in the near-surface layers. 
The extent of the thermal modifications sustained during the dynamical evolution of Centaurs is only starting to be 
quantified. 
As a result of the complex and varied dynamical pathways from the outer solar system, it is impossible to constrain the exact orbital evolution that any given Centaur has followed beyond the last interaction with Jupiter, nor is it possible to know whether it is on its first pass to the inner solar system. Constraining the degree of thermal processing sustained by each individual Centaur is, consequently, extremely challenging. 
We expect that distinct individual orbital pathways should result in significant differences in the internal composition and structures when Centaurs are observed. As water ice does not efficiently sublimate in the giant planet region, erosion is limited and chemical differentiation could occur down to several hundred meters \citep[depending on thermo-physical properties,][]{guilbert2016}.
An effort to provide more realistic constraints on the degree of thermal processing sustained during the Centaur phase has recently been made, by coupling both the thermal and the orbital evolution of Centaurs and JFCs. Indeed, \citet{gkotsinas2022} showed that as a result of the stochastic nature of the dynamical trajectories, each Centaur may experience multiple, long-lasting heating episodes, leading to chemical alteration of their upper layers, down to several hundred meters, allowing for a substantial depletion of their super-volatile content (Fig.~\ref{fig:orbit-evolution}).
\new{T}he relative lack of compositional signatures in the coma of active Centaurs is impeding further progress, a status that is currently changing with JWST observations of these objects.

Recently, attempts to identify Centaurs on the edge of transitioning to a JFC-like orbit have been made. These would allow for an efficient and detailed investigation of how dynamical and thermal evolution alters comet nuclei before they become JFCs \citep{sarid2019,steckloff2020,kareta2021,seligman2021}, especially as potential targets for future space missions. However, \citet{guilbert2023gateway} showed that most objects found near the orbit of Jupiter are statistically more altered than the rest of the Centaur population, due to thermal processing sustained in previous stages of dynamical evolution \new{since} these objects might have already been JFCs in the past. 
In addition to long-term effects discussed above, seasonal and diurnal patterns of activity should act on Centaurs as they do for JFCs. For these smaller timescales, the orientation of the spin axis is key (see \citealt{Fernandez2025}), as it can cause insolation patterns on the surface that trigger thermal behavior relatively unique to each active object.

\subsection{Impacts and Cratering}
\label{subsec:cratering}

Cratering is a prominent shaping process in the solar system, and every object experiences impacts during its lifetime. \new{Throughout the solar system, craters are used to date the surfaces of objects, to derive the size distribution of the impactors and to constrain the dyanmical history of different populations. However, identifying craters on the surfaces of active bodies is complicated by activity-driven erosion. Thus, there are few unambigous crater detections on the surfaces of comets visited by spacecraft besides the crater produced by Deep Impact on 9P/Tempel~1 \citep[see \final{review by}][for details]{Pajola2023}. 
In this section,} 
we consider the limited time period during which an object is classified as a Centaur. As the population of Centaurs is not well constrained, it is difficult to establish reliable probabilities of collisions within the population.
The best alternative is to consider other, better calibrated sources of impactors. 

In the outer solar system, the main reservoir of  projectiles is the population of eccentric comets, which are known to collide with giant planets and their moons. Impacts on Jupiter are routinely observed and amount to about one event per year for 10~m diameter projectiles \citep{Hueso2010}. A recent review of cratering records on the moons of giant planets, combined with dynamical models of the solar system, allows an estimate of the collision rate for small objects impacting the moons \citep{Nesvorny2023}. The impact rate and velocity are variable and depend on both heliocentric and planetocentric distances. For projectiles larger than 1~km, the average time between collisions varies from 2.7~Myrs to 42~Myrs, which is comparable to the dynamical lifetime of Centaurs. 
\final{At the same heliocentric distance, collisions on objects unbound from planets are less frequent as the impact rate on a moon is magnified by the focusing presence of its host planet.}
We can therefore conclude that larger impacts, which would reshape or destroy Centaurs, are very unlikely to occur in the dynamical lifetime of these objects. In other words, the current population is not shaped by impacts.
Smaller scale collisions are possible, and future missions to Centaurs may observe craters in the surface of these objects, but most impacts will have occurred in earlier dynamical phases  
(see Section 4.5.2 in \citealt{Fernandez2025}).

\subsection{Close encounters}

Due to gravitational interactions with the giant planets, Centaurs' orbits are chaotic and an object's orbital elements can vary \new{considerably}
during its dynamical lifetime. One consequence is that Centaurs \new{may} experience close encounters ($<$~5~planet radii) with the giant planets. \new{Earlier  
dynamical calculations for the  observed population showed that at least 10\% of the known Centaurs will encounter a giant planet within its Roche limit \citep{Tiscareno2003, Hyodo2016}.}
\new{However, more recent works suggest that close encounters with the giant planets are very rare \citep[e.g.,][]{Araujo2016,Safrit2021}. The orbital integration performed by \cite{DiSisto2025}{} suggests that only $\sim$2\% of the Centaur population have an encounter within the Roche limit of a giant planet in the age of the solar system. } 

Such a close encounter 
\new{would induce}
strong tidal forces in the Centaur, which can have significant effects on its surface morphology and global shape. 
\new{During a close encounter, the gravitational pull of the planet creates a differential force that may stretch the Centaur along an axis aligned with the planet, or at the very least lower the effective gravity in areas facing toward and away from the planet. As a consequence, the local slope of the terrain\final{,} measured as the angle between the surface normal vector and the effective gravity (self gravity + tidal force + centrifugal force)\final{,} can be modified. Even if the body is not stretched by tides, the direction of the effective gravitational acceleration will change during the encounter, effectively raising or lowering the local slopes by a few degrees. This effect has been studied and modeled for close encounters of Near Earth Objects with our planet, for instance asteroid (99942) Apophis \citep{Kim2023}}.
While modeling this process for Centaurs remain\new{s} challenging given our limited knowledge of the mechanical properties of these objects, simulations of tidal effects on Near Earth asteroids show that tidal forces \final{could be contributing to Centaur evolution}
\citep{Kim2023}. 
At their most extreme end, tidal forces can lead to a complete shearing of the body and catastrophic disruption. A prominent example of such an event is comet D/1993 F2 (Shoemaker-Levy 9), which was disrupted by a close encounter with Jupiter in July 1992 and eventually collided with the planet two years later \citep{Nakano1993}. 

Recent observations have revealed the existence of icy rings around two large Centaurs: \new{(}10199\new{)} Chariklo \citep{Braga2014} and (2060)~95P/Chiron \citep{Ortiz2015}.
\new{Understanding the mechanisms which could lead to the formation of Centaur rings is still in its early phase and a few scenarios which involve a close encounter with a giant planet are under consideration (see \sickafoose{} for a review). For instance, one of the proposed hypotheses}
\citep{Hyodo2016} is that a close encounter with a giant planet could have stripped the icy mantle of 
\new{a Centaur.}
The resulting debris would reassemble within the Roche limit of the remaining core, forming a ring. 

While the mechanisms responsible for ring formation remain poorly constrained (see \sickafoose{}), the first attempts 
to investigate \new{the possible connection between close encounters and} ring survivability provide 
instructive results. \cite{Araujo2016,Araujo2018} performed numerical simulations to study the encounter probabilities of Chariklo and Chiron \new{with the giant planets}. Their work identified that tidal forces are less efficient at disrupting the ring systems of more massive bodies, as well as of objects with high-inclination and low-eccentricity orbits. \new{Recently, the possibility that a close encounter with another small body could disturb a Centaur's ring system was considered by \citet{Ikeya2024}. They found that while it is possible for an encounter with a similar-sized body to perturb an object's rings, such disruptive close encounters are extremely rare and unlikely to occur in 4 Gyr. }


%


\section{Evidence for evolutionary processes}
\label{sec:evidence}

The overview in Section \ref{sec:processes} clearly exposes the limitations of our current knowledge regarding the evolutionary processes in the Centaur region. In this section, we focus on key observational and modeling results. Our goal is to review how they have shaped the present understanding of evolution mechanisms and to highlight some intriguing manifestations of Centaur evolution worthy of future investigation.

%


\subsection{Activity}

A considerable fraction (estimated at 10-15\%) of the known Centaurs display comet-like activity \citep{Jewitt2009,Bauer2013}, which is driven by the dynamical and consequently thermal evolution of the object (see Section \ref{sec:proc-thermal}). Hypothetically, Centaur activity opens the possibility to characterize the volatile content of the population. However, we know from a modeling point of view that single measurements of gas production rates in the coma of an active object \new{can be misleading in relation to the abundance of volatile species inside that object's nucleus} 
\citep[e.g.,][]{benkhoff1995,prialnik2006}.
This diagnosis can be extended to Centaurs, which evolve in a region of the solar system where critical phase transitions for a variety of volatile species are met, so that relative abundances in their 
\new{coma do not necessarily depend only on heliocentric distance,}
but also \new{on} seasonal and possibly diurnal cycles \citep{Guilbert2023-cometsiii}. 
Centaur nuclei can furthermore be heterogeneous in composition \citep[\final{as recently evidenced by the JWST observations of 29P/Schwassmann-Wachmann~1 by}][]{Faggi2024}, either because they formed heterogeneous, or because the layer of material contributing to the observed activity \new{has inevitably developed} a non-uniform structure and composition as a result of prior thermal evolution (see \citealt{Guilbert2023-cometsiii} and Section \ref{sec:proc-thermal} above).

Detecting any volatile signature in the \new{comae} of active Centaurs has proven
\final{elusive}
up to now \new{because the main process by which gas is 
\final{observable} is through its fluorescence excitation \citep[cf.\ reviews by][]{Biver2022,Bodewits2022}. The combination of large heliocentric and geocentric distances combined with very low levels of activity makes the direct detection of volatiles extremely challenging}. For instance,
\final{no Centaur has been detected to be active}
beyond $\sim$12~au \citep[e.g.,][]{li2020}, and no 
\final{definitive}
detection of gaseous CO \new{was} made in the pre-JWST era, except for 29P \new{(see \citealt{Kareta2025} for a detailed review of 29P's activity characterization)}. 
Marginal detections of CO were reported for Chiron \citep{womack1999,womack2017},
and \new{(60558)} 174P/Echeclus \citep{wierzchos2017}. 
These observations are consistent with the activity of Centaurs not being primarily driven by the sublimation of CO \citep{drahus2017}.  
\new{A few fragment (CN) and ionic (CO$^+$, N$_2$$^+$) species have been detected in 29P \citep{Cochran1991,Korsun2008} but \new{we highlight that this \final{object} has exceptionally high activity.}}
\final{The recent detections of gas signatures in the coma of several active Centaurs by JWST are game changers for understanding how the evolutionary processes work in the giant planet region, and to pinpoint the activity mechanisms. For instance, CO$_2$ was detected in the coma of active Centaur 39P/Oterma \citep{Pinto2023}, and both CO and CO$_2$ are detected in the coma of 29P \citep{Faggi2024}. Emission features from CO$_2$ and CH$_4$ are detected in the spectrum of 95P/Chiron, suggesting a substantive level of activity, although the coma itself is not directly detected \citep[][in revision]{Pinilla2024}.
Only a careful analysis of this object's phase curve provides hints for a new epoch of activity \citep{dobson2024}. We point to \cite{FV2025}, \cite{Bauer2025}, and \cite{Mandt2025}{} for further discussion. }

Patterns of activity can yield information about the surficial layer of material, as a number of Centaurs have been reported \new{to be active} 
several times. For instance, the activity of 29P has been documented for several decades, as this Centaur appears to be nearly always active with dramatic outbursts reported on a regular basis 
\new{(see \woodney{} and \kareta{} for comprehensive reviews)}.
Chiron has been reported active at least twice since its discovery: a first period of activity lasted $\sim$6~yrs when the object was between 9 and 12~au \citep[see review by][]{womack2017}, and \final{a more recent epoch of} outbursts in 2021 \final{near aphelion} at $\sim$18.8~au \citep{dobson2021,Ortiz2023}\final{, with the activity from 2021 apparently continuing through at least early 2023 \citep{dobson2024}}. 
A puzzling aspect of Chiron is that basic characteristics such as its activity, near-infrared spectrum, brightness, and colors vary with no clear correlation with heliocentric distance \citep[][and references therein]{womack2017}. The detection of a ring system around Chiron has provided 
\new{new insights into}
these unusual changes \citep{Ortiz2015,ruprecht2015}, however, these cannot explain the occasional outbursts and development of a coma. 
Another Centaur, Echeclus is
\new{characterized} by repeated outbursts of activity, with at least five events reported to date \citep[see][and references therein]{rousselot2021}. 
\final{Each outburst could have its own origin (e.g., fragmentation events, collisions), however outbursts could also reflect internal heterogeneities as described by \citet{rosenberg2010Icar..209..753R}.}
\final{Indeed, those authors show that} non-uniform structures (both in composition and thermal properties) can cause erratic behaviours with outbursts of diverse intensities and durations being triggered at any point of the orbit, even at large heliocentric distances, with gas production rates that can even exceed those expected at perihelion. 
\new{Continuous} monitoring of active Centaurs, coupled with gas detections in their coma, are therefore necessary to provide crucial insights into the mechanisms driving the activity evolution of this population.




\subsection{Physical properties}

\subsubsection{Size and Shape}
\label{sec:size_and_shape}

The Centaur size distribution and how it compares to other populations is discussed in \cite{Fernandez2025}. That chapter mainly focuses on the extent to which the size distribution of today's Centaurs is informative about the original planetesimal properties. Here, we pose a different question: do the sizes of Centaurs contain any indications of significant evolutionary processing during the Centaur phase? In Section \ref{sec:processes} we discussed tidal interactions and impacts. In addition, the sizes and shapes of Centaurs can also be modified by outbursts, rotationally-driven instabilities and splitting. While each of these processes can influence the fate of an individual object, it is unclear whether a large fraction of the population is subject to significant size modifications during their lifetime in the Centaur region.

Answering this question is unlikely to be possible with current data given the uncertainties in the size distributions of Centaurs and the other populations from the main evolutionary pathway (excited TNOs and JFCs, see \citealt{Fernandez2025}). A major challenge is in detecting 
\final{similarly}
sized objects for comparison\final{; most JFCs are much smaller than the known Centaurs, while most TNOs are larger}. Future studies, comparing the small members of each population are essential (see Section \ref{sec:outlook}).

Considering the global size distribution 
of potential impactors, we established in section \ref{subsec:cratering} that there is no evidence for collisional evolution playing a major role in shaping the current size distributions of objects while in a Centaur phase. Other effects, like sublimation-driven activity, are even less likely to modify this size distribution\final{. We have so far not observed outbursts large enough to indicate a total disruption of a Centaur (see \bauer{}) and observations indicate that dust-production rate similar to that observed for 29P \citep[e.g.][]{Stansberry2004} would not be efficient at modifying the size of a Centaur like 29P within the typical dynamical lifetime in this region.} 
It is therefore likely that the current size distribution is a mix of the size distributions inherited from the parent groups of the observed objects, and does not inform us on processes taking place in the Centaur phase. 

To study Centaur evolution, it is perhaps more \final{informative}
to consider the global shape rather than size, especially when it departs from a simple ellipsoid. Of particular interest are objects like binaries (either in contact or separated), or Centaurs with rings, as these features can be the signatures of recent large scale reshaping events and would then bring insights on the physical and mechanical response of the objects.

\new{Radar and spacecraft observations have revealed that bilobate shapes could be widespread among comet nuclei. Out of the seven comets whose shapes have been determined, five appear bilobate \citep{Keller2015,Hirabayashi2016,Safrit2021}}.
Binaries are \new{also} common in the TNO region \citep{Stephens2006,Noll2020}, see also the discussion in \citet{Bernardinelli2023}. However, the absolute percentage of binaries among TNOs transitioning to JFCs is unknown,  
as is their formation mechanism, but we know that bilobate small bodies may be formed at large heliocentric distances, as evidenced by the images of Arrokoth returned by NASA's New Horizons mission \citep{Spencer2020}. Two other TNOs crossing the Centaur region, are also known to be binaries: (42355) Typhon and Echidna \citep{Noll2006} and (65489) Ceto and Phorcys \citep{Grundy2007}. Statistical analysis of the available observations sets a $3\sigma$ limit $<8\%$ for the fraction of binaries in the Centaur population \citep{li2020}.

Although the data are too sparse to conclude that there is an excess of bilobate objects in the JFC population, compared to their progenitor groups, it is worth considering whether there could exist a mechanism to transform singular objects into binaries before they become JFCs. A promising candidate would be rotational instability triggered by sublimation torques \citep{Safrit2021}, but it remains speculative until the detection of a bilobate and/or fast-rotating Centaur. We discuss rotational stability in more details in the next section.

\subsubsection{Spin state, density and strength}

\begin{figure}[t!]
\begin{center}
\includegraphics[width=.8\textwidth]{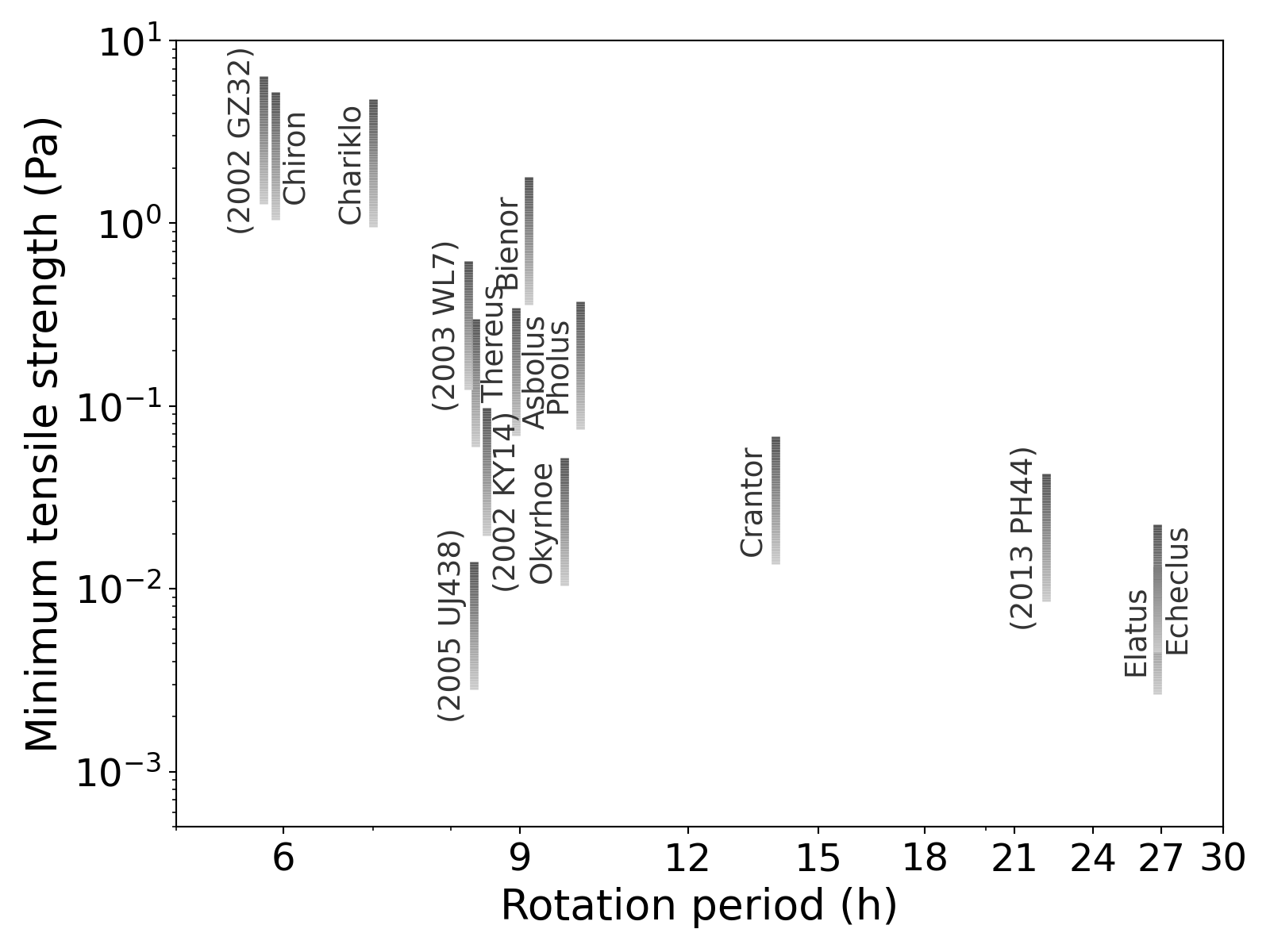}
\end{center}
\caption{Minimum tensile strength required for Centaurs to spin at their observed rotation period without breaking apart. For each object, the line gradient represents a range of possible densities (from light to dark: 500~kg m$^{-3}$ to 2500~kg m$^{-3}$). To preserve figure readability, Centaur 2013 XZ8 is not displayed ($P~=~88$~h, $Y_t < 10^{-3}$~Pa). } 
\label{fig:strengh_vs_rotation}
\end{figure}

As has been discussed in detail for JFCs \citep[e.g.,][]{Knight2023}\final{, Long period comets \citep[e.g.,][]{Jewitt2022_destruction}}, and Centaurs (\citealt{Fernandez2025}), activity-driven torques can change an objects' spin state significantly and even lead to rotational instability or breakup. 
The rotational period of small bodies can thus be used to estimate a lower bound for their internal cohesion. Beyond a critical spin rate, the resulting tensile stresses overcome gravity and cohesion, leading to a catastrophic disruption of the object. \new{For bodies with extremely low cohesion (e.g., as observed with asteroid rubble piles), balancing gravitational and centrifugal accelerations on the surface gives a typical critical period of 2.2~h: $P_0 = \sqrt{\frac{3\pi}{G\rho}}$ with $\rho$ the density (typically $2300~$kg~m$^{-3}$), and $G$ the gravitational constant. We do not know the density of Centaurs but it \final{is} reasonable to consider the range $[500 - 2500]$~kg~m$^{-3}$, i.e., comet-like to Trojan-like. For the lowest density, the critical period becomes 3.3 h.}
Any object spinning at a faster rate will start shedding material from its surface, and eventually be destroyed. This is a well accepted hypothesis to explain the dearth of fast rotating asteroids at sizes larger than a couple hundred meters \citep[e.g.,][]{Pravec2000}. \new{We note that this preceding discussion is only for gravity-dominated objects; data are insufficient to speculate on strength-dominated objects.}

When considering cohesion and inner structure, calculating the spin limit requires more complex models which properly account for the distribution of stresses  \citep[e.g.,][]{Scheeres2018, Li2021}. \new{However, one can approximate the maximum rotation rate within an order of magnitude with $\omega_c = \sqrt{\frac{2Y_t}{\rho d^2}}$ where $Y_t$ is the tensile strength and $d$ is the object diameter \citep{Sanchez2014, Safrit2021}.}
The critical period is then simply $P_c = 2\pi / \omega_c$.

Very little is known about the mechanical properties and internal structure of Centaurs, and we only have measurements of size and rotation period for 16 objects (\citealt{Fernandez2025}). But we can nonetheless calculate what would be the minimum tensile strength required by Centaurs to withstand their current observed spin rate. Although we do not have measurements for the density, it is reasonable to assume that it lies somewhere in between typical densities of comets (500~kg m$^{-3}$) and Trojans (2500~kg m$^{-3}$). Using this range, we calculated the lower bound for the tensile strength of all objects (Fig.~\ref{fig:strengh_vs_rotation}). \new{We find that even the fastest observed Centaurs could sustain their current rotation rates with as little as 1 Pa of cohesion, which is lower than the typical tensile strength of objects at comparable size and heliocentric distances, see discussion in \cite{Biele2022}}. The actual tensile strength of Centaurs is almost certainly larger than this boundary. At comparable size range and heliocentric distances, comets would be the weakest objects \new{known} and they are stronger than this limit (see Fig.~8 in \citealt{Biele2022}).

It is therefore unlikely that any of the current Centaurs with known rotation period\new{s} will break up due to their spin. Future surveys, however, may reveal fast rotators in the Centaur population, which would bring significant insight on the mechanical properties of these objects.



\label{sec:evidence-surface}

\subsection{Surface properties}
\label{sec:surface_properties}

The surface properties of Centaurs, and especially their colors have been used extensively to characterize the population. This book includes chapters on the surface chemical composition (\citealt{Peixinho2025}{}) and on the recent JWST spectra beginning to revolutionize our understanding of Centaur surfaces (\fv{}). We refer the reader to these papers for details, and here we focus on selected observational evidence that can provide insight into the evolutionary processes introduced in Section \ref{sec:processes}.

\subsubsection{Color index}
\label{sec:evidence-colors}

The most widely characterized Centaur property is undoubtedly optical photometric color. Early studies of the visible spectra of TNOs, Centaurs, and JFCs revealed predominantly featureless surface spectra\new{. The reflectance properties of Centaurs can thus be reliably described using parameters such as the normalized reflectivity gradient between two end points, $S'$ \citep{AHearn1984}, and color indices, which measure the photometric difference between the object's magnitude in standard broadband filters (e.g., $B-R$, $g'-r'$, etc.).} 
Centaurs and TNOs display a large diversity of colors, spanning from solar-like neutral or grey surfaces with $B-R~\sim~1.0$~mag to very red colors, $B-R~\sim~2.0$~mag.

A remarkable feature of the color distribution of small bodies in the outer solar system is the existence of very red objects. Most works in the literature use a limit of $B-R~>~1.6$~mag (which corresponds to $g'-r'~>~ 0.75$~mag in the SDSS system) or spectral gradient $S'\geq~25\%$~per~100~nm to define the so-called `ultrared' population \citep[e.g.][]{Jewitt2002,Lacerda2014,Schwamb2019}. A series of multiband photometric studies have found evidence that the color distribution of Centaurs is bimodal \citep[see][and references therein]{Peixinho2015,Tegler2016}. Even though the color bimodality is most pronounced in Centaurs, it has also been found in dynamically excited TNOs \citep[e.g.,][and references therein]{Peixinho2015,Tegler2016,Wong2017,Marsset2019}. The two surface-color types are also characterized by different geometric albedos. The very red objects have more reflective surfaces with geometric albedos $\sim$0.15 while the more neutral objects have small geometric albedos of $\sim$0.05 \citep[e.g.,][]{Lacerda2014}. 

\new{A} few hypotheses have been \new{proposed} (see \citealt{Peixinho2025} for a review), \new{but} the underlying reason for the existence of the two surface types remains uncertain. Likewise, the nature of the ultrared matter remains elusive. The current preval\new{ing} hypothesis is that ultrared objects have a thin ($\sim$meter thick) shell of complex organic matter resulting from prolonged space weathering in the outer solar system \citep[e.g.,][]{Thompson1987,Cruikshank1998,Jewitt2002,DalleOre2015}. The existence of very red objects is thus \new{thought} to reflect compositional differences in the PKB.

Interestingly, red Centaurs have a smaller orbital inclination angle distribution than gray Centaurs \citep{Tegler2016}. This color-inclination relationship is considered to be inherited from the color-inclination correlation observed throughout the Kuiper Belt \citep[e.g.,][]{Marsset2019}.
In the case of Centaurs, however, the color distribution is hypothesized to also be modified by ongoing evolution in the giant planet region. One of the most remarkable observations in favor of this argument is the finding that ultrared objects have not been discovered at heliocentric distance below $\sim$10~au \citep{Jewitt2009,Jewitt2015}\footnote{
\new{Although} 
there is consensus in the literature that ultrared matter disappears below $\sim$10~au, we note that \cite{Lamy2009} identified a few JFCs with nucleus color indices $V-R~>~0.63$~mag which corresponds to a normalized spectral gradient larger than 20\%~per~100~nm and could be classified as ultrared. We caution that without more sensitive photometric observations and a larger band coverage, this remains tentative evidence.}. Moreover, this location coincides with the heliocentric distance where Centaur activity begins, which motivated the hypothesis that the onset of activity leads to the disappearance of ultrared matter \citep{Jewitt2002,Jewitt2009,Jewitt2015}. 
This hypothesis is supported by further evidence in the orbital and color distributions of Centaurs. Dynamical simulations of Centaur orbits show that neutral Centaurs have longer dynamical lifetimes than red ones \citep{Melita2012} and that inactive Centaurs have longer dynamical lifetimes than active ones \citep{Fernandez2018}. These finding\new{s} can be explained by postulating that a sufficiently long dynamical lifetime orbiting among the giant planets allows activity onset and leads to loss of the very red surface colors.

To explain the disappearance of \new{ultrared} matter in the inner solar system, \cite{Jewitt2002} explored mechanisms in which the onset of activity would destroy or occlude the very red colors of Centaur surfaces. Their work, followed by further investigations \citep[see][]{Jewitt2015} established that blanketing of the surface by fallback of material ejected due to outgassing is a plausible mechanism which can explain the rapid disappearance of ultrared colors with the onset of activity (see also the discussion on surface mobilization in  Section \ref{sec:evidence-surface-mobil}). \new{This airfall process was, in fact, \final{observed}
on comet 67P\new{/Churyumov-Gerasimenko} by Rosetta \citep{Thomas2015,Keller2017,Davidsson2021b}, and it was associated with color changes on the surface of the comet \citep[see][]{Filacchione2020}. However, because of our limited understanding of Centaur activity, it is unclear whether parallels could be made with processes taking place in the giant planet region. It is also important to note that the global optical color change of 67P's nucleus could not be determined precisely because of uncertainties in the variation of the phase-reddening curve (defined below) and can only be approximated to changing from $V-R\sim0.52$ mag in August 2015 to $\sim$0.48 mag during perihelion (Fornasier, private communication).}

Even though th\new{e} hypothesis of Centaur surface evolution has been widely accepted, it is now being re-examined after the discovery of the first active red \new{C}entaur \new{(523676)} 2013~UL10 \citep{Mazzotta2018}. To explain the existence of this object, and also invoking the dynamical trend that all neutral Centaurs (inactive and active) have orbits with smaller perihelia than red Centaurs, \cite{Wong2019} proposed that thermal processing can be sufficient for the disappearance of ultrared matter without the need for activity onset. Alternatively, \citet{Ortiz2015} speculated that since the two Centaurs with ring detection (Chariklo and Chiron) are both neutral in color, the optical colors of Centaurs can be related to the presence of rings. Even though the evidence supporting this claim is sca\new{nt}, as more occultations are observed \new{(potentially detecting or setting constraints on the existence of rings; see \sickafoose)}, they might shed light onto the possible connections between surface evolution, activity and rings/debris around Centaurs.

Another noteworthy characteristic of the color bimodality of Centaurs is its dependence on size. The color bimodality is evident only among small \new{(diameter $<120$~km)} Centaurs 
\citep{Duffard2014}. 
\final{However, the dependence of color on size has}
 not been 
 the focus of Centaur investigations and it remain\new{s} unclear whether it is a consequence of Centaur evolution, or whether it can be attributed to the properties of the source populations of Centaurs and/or observational biases. Notably, recent observations have indicated that very red Neptune Trojans also show a possible trend with absolute magnitude \citep{Markwardt2023}. 
 \final{It will remain difficult to investigate the existence of relationships between Centaurs' size and surface properties}
 unless significant efforts are directed into thermal infrared surveys of outer solar system populations (see Section \ref{sec:outlook}).


\subsubsection{Surface spectra}

Before JWST, 
\new{near infrared}
spectra had  been obtained for a small sample of Centaurs as parts of surveys primarily targeting TNOs \citep{Barkume2008,Guilbert2009,Barucci2011,Brown2012}. Recent and future observations with 
\new{the JWST Near Infrared Spectrograph (NIRSpec)}
will allow for a more detailed understanding of 
\final{which surface properties} 
Centaurs inherit from their TNO parents, and how their heightened thermal processing affects their surface properties (see \fv{} for further discussion). \new{JWST is also opening the possibility to compare Centaur surfaces to those of more evolved objects, especially JFCs \citep[see][]{Kelley2016}.}

\new{Some of the most intriguing JWST results 
\final{on the links between Centaur and TNO surface properties}
have so far come from the large JWST program \final{``}Discovering the composition of the trans-Neptunian objects, icy embryos for planet formation\final{"} (DiSCo-TNOs, Program ID: 2418)}.
\new{Using these data,} three spectral types have been identified in the trans-Neptunian region: (1) a first group with spectra dominated by water ice, (2) a second group with spectra dominated by CO$_2$ and CO ices, and (3) a third group with spectra dominated by carbon bearing ices and organic species \citep{Pinilla2025,DePra2024} 
The first and third groups 
\new{(dominated by water ice, and carbon-bearing ices and organics, respectively)} 
are also detected in the sample of Centaurs targeted by JWST \citep[see][]{Licandro2025}: some grey Centaurs display water-dominated spectra, while some red Centaurs display C-bearing and organic species. We note that both spectral types, when observed on Centaurs, appear slightly depleted in volatile species as compared to TNOs, with shallower absorption bands on average testifying to lower amounts of the corresponding ice at the surface. 
As of now, the second group (CO$_2$- and CO-ice dominated) appears absent in the giant planets region, which might
\new{suggest}
that increased surface temperatures lead to the sublimation of these volatile species.

Interestingly, a distinct spectral group is detected in the Centaur sample, which does not correspond to any parent spectral type in the trans-Neptunian region \citep[see ``misfits'' in][]{Licandro2025}:
these objects display both red and grey optical colors. Whether these coincide with a thermally-processed version of one or several of the TNO parent spectral types remains to be resolved.

Additional insights on the compositional differences between objects with different surface types and on the processing responsible for surface evolution at different environments in the solar system can be obtained from comparisons with other populations. The first steps in that direction are already being taken in the analysis of recent JWST spectra of Jupiter Trojans \citep{Wong2024} and Neptune Trojans \citep{Markwardt2023_JWST}.

\subsubsection{Other spectrophotometric properties}
\label{sec:evidence-photom}

While the surface colors of a large fraction of the known Centaurs have been measured, the other spectrophotometric properties of the surfaces remain poorly constrained. There are two main limitations in characterizing the reflectance properties of Centaurs from the Earth: the faintness of the targets, and the small phase-angle (the Sun-target-observer angle) range, limited to a maximum of \new{$\sim$10$^\circ$}.

Geometric albedo is defined as the ratio of the reflected light to the incident light at zero phase angle. It has been determined for more than 40 Centaurs \citep{Mueller2020}. This relatively large sample has been achieved mainly as a result of large thermal-IR observing programs of TNOs with Spitzer \citep[see][]{Lisse2020} and Herschel \citep{Mueller2009}. Deriving the geometric albedos is possible when the optical brightness (absolute magnitude) of the object is compared to the size estimate derived from thermal IR observations. The latter relies on the use of thermal models (e.g., NEATM, \citealt{Harris1998}) and assumptions about the surface properties of Centaurs (via the emissivity parameter of the models). 

Compared to JFCs, \new{whose geometric albedos} range from 0.01 to 0.06 \citep{Knight2023}, the albedos of Centaurs span a much larger range from $\sim$0.04 to $\sim$0.25 \citep{Mueller2020}. Notably, Centaur albedos are neither correlated with their orbital parameters, nor with the objects' diameters (see \citealt{Duffard2014} and \citealt{Fernandez2025}). The only pronounced trend identified in the literature is that red Centaurs have a larger median albedo \citep{Duffard2014,Lacerda2014,Tegler2008,Tegler2016,Romanishin2018}. Additionally, the albedos of grey Centaurs are found to be similar to the albedos of JFCs \citep{Duffard2014,Mueller2020,Knight2023} and Jupiter Trojans \citep{Romanishin2018}. However, these limited clues are insufficient to derive any meaningful constraints on the evolutionary processes proposed for the Centaur region.

As can be seen in Table~7.1 in \citet{Mueller2020} and Fig.~\new{4}.3 in \cite{Fernandez2025}, the uncertainties of the albedo estimates are very large. It is possible to derive more precise estimates of the object size and therefore of the geometric albedo in the cases of occultation observations (see \sickafoose{}). 
Additionally, accurate albedo determinations 
\new{can be improved by more}
precise estimates of the absolute magnitude (\new{defined as} the \new{apparent} magnitude the object would have if it were placed at 1 au from the Sun and the Earth and at zero phase angle.
Therefore, the \new{calculation of an object's} geometric albedo depends also on the phase-function assumed.

The phase function describes the variation of the reflectance with phase angle.  Typically, the small phase angles at which Centaurs are observable are characterized by an opposition effect or a steep increase of the reflectance \citep{Gehrels1956,Belskaya2000}. However, instead of the broad and sharp opposition effect characteristic of asteroids with moderate and large albedos \citep{Belskaya2000}, Centaurs and TNOs display a shallow linear slope down to $0.1^{\circ}-0.2^\circ$ and a narrow, weakly pronounced opposition effect \citep{Belskaya2000,Belskaya2006}. Most Centaur phase functions are observed to be linear and are described by \new{a linear slope, $\beta$, known as the} 
phase coefficient \citep[\new{see}][]{belskaya2008}. This behavior is similar to that of comets \citep[see][]{Knight2023} and Jupiter Trojans \citep{Shevchenko2012}.  

\begin{figure}[t]
\begin{center}
\includegraphics[width=0.7\textwidth]{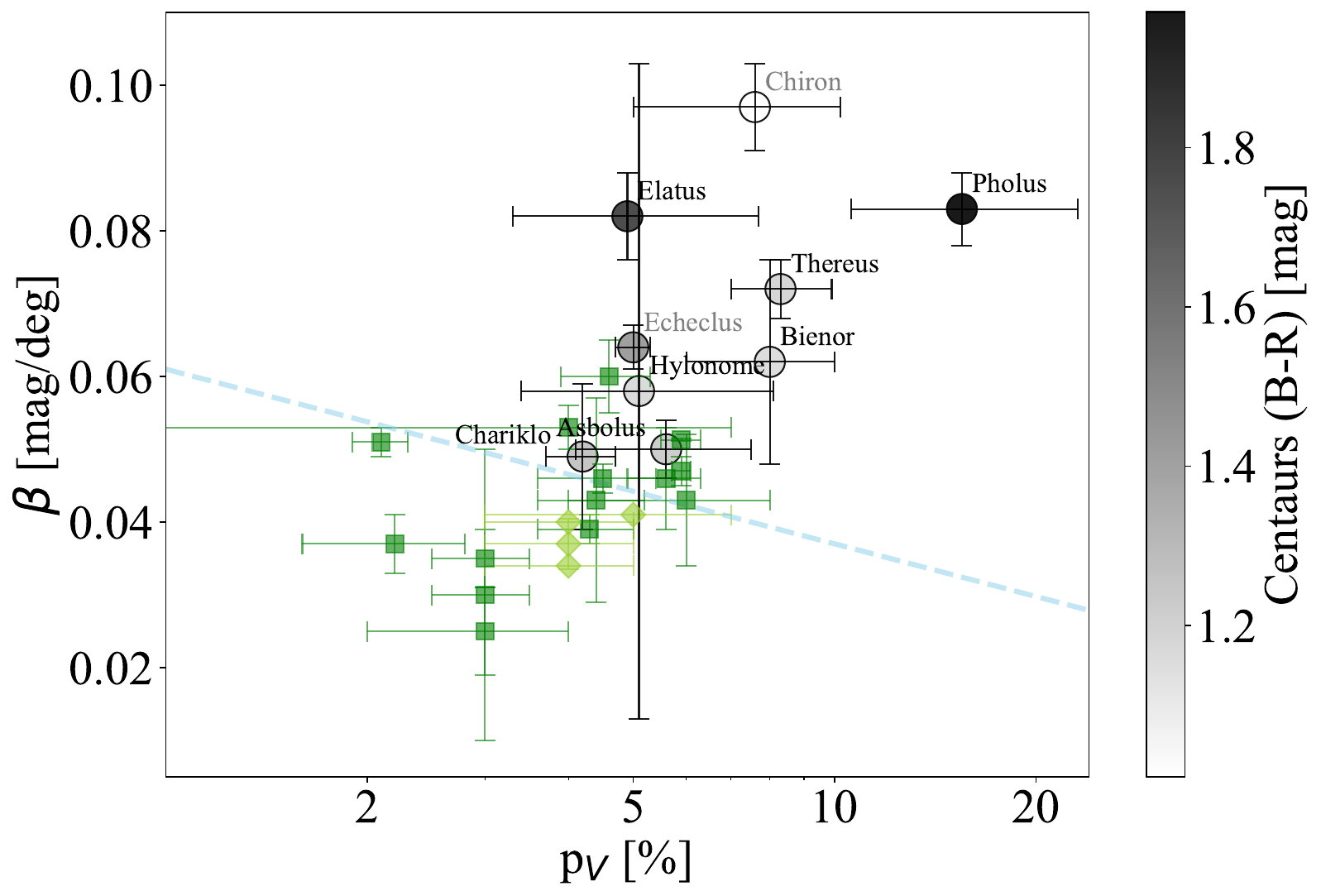}
\end{center}
\caption{Apparent correlation between the phase coefficient ($\beta$) and geometric albedo ($p_\mathrm{V}$) of JFCs (green squares) and other comets (light green diamonds) from \cite{Knight2023}. Centaurs are overplotted as circles with colors corresponding to their $B-R$ color index. The phase coefficients of Chiron and Echeclus were derived from observations when they were weakly active \citep{Dobson2023}. 
\new{With the exception of the very-red (Elatus and Pholus) and the active (Chiron and Echeclus) Centaurs, other Centaurs}
appear to follow the correlation of increasing phase coefficient with geometric albedo identified by \cite{Kokotanekova2018} and do not agree with the phase-coefficient--albedo correlation found for Main-Belt asteroids by \cite{Belskaya2000}, plotted as a dashed line.
}
\label{fig:surface-prop}
\end{figure}

Even with the limited observational evidence on Centaur phase functions, it is interesting to make a comparison between Centaurs and JFCs. \cite{Kokotanekova2018} suggested that the 
\final{possible correlation of smaller phase coefficient for decreasing albedos}
may be related to the level of sublimation-driven erosion on their 
\final{surfaces, which is hypothesized to decrease the phase coeffient and albedo of the comets with time}
(see also section \ref{sec:surface_morphology}). In Fig.~\ref{fig:surface-prop} we plot the Centaurs \new{and JFCs} with known phase coefficients and albedos. We note that the distributions of Centaurs in this plot is in agreement with the prediction that objects with less evolved surfaces should have larger albedos and phase function slopes.
However, this plot is based on very limited data and should not be overinterpreted before further photometric data can be used to derive more reliable phase-function estimates for a larger sample of Centaurs and JFCs.


Finally, a review of the surface properties of Centaurs would not be complete without mentioning polarization. Polarization observations of outer solar system objects are challenging due to the targets' faintness and limited phase angle range  \citep{Belskaya2015,Belskaya2019}. To date, only the polarization of a handful of the brightest Centaurs has been observed and these observations with their possible implications have already been reviewed by \citet{Belskaya2015} and \citet{Belskaya2019}. Within the narrow phase-angle range ($0.5^{\circ}-4.4^{\circ}$), Centaurs display a great diversity of their polarization properties which has been attributed to different abundance of ices on their surface \cite[see][]{Belskaya2015}. Polarization is therefore showing promise to serve as a diagnostic of the surface evolution of Centaurs but this would only be possible if the number of Centaurs observed increases, polarization is measured in multiple bands, and advances in numerical and laboratory work are made.

\subsection{Surface morphology}
\label{sec:surface_morphology}

\subsubsection{Evolution of surface topography}

A number of recent studies have established a possible link between features observed at the surface of JFCs (like topography or roughness), and the evolution state of that surface.
\new{With the detection of numerous pits at the surface of 9P, \citet{belton2008,belton2013} tried to link the occurrence of mini-outbursts with the origin of such morphologic features. Later, through a statistical analysis} of the distribution of large-scale topographic features imaged on 67P and other comet nuclei imaged by spacecraft missions, \citet{Vincent2017} identified that the height of cliffs correlates with the rate of surface erosion. 
This \new{suggested} that topography could be used to trace a surface's erosional history, \new{which is a conclusion independently derived by \citet{steckloff2018} using a theoretical approach.}

Inspired by \new{these results}, \citet{Kokotanekova2018} proposed that the correlation they had previously reported in the phase function of 14 JFCs \citep{Kokotanekova2017} could be related to surface topography. In this framework, rough and more primitive surfaces, characterized by a stronger phase darkening, would gradually evolve toward smoother terrains with smaller phase coefficients. This is supported by theoretical calculations showing that pure geometric effects stemming from sharp surface topography represents a non-negligible contribution to the phase darkening \citep{Vincent2019}.
A physical framework can be provided by establishing quantitive trends from the thermal processing of cometary surfaces. 
Thermal evolution calculations suggest that cometary activity tends to erase sharp morphological features, which become wider and shallower over time \citep{Benseguane2022,guilbert2023pits}. 
Consequently, pits observed at the surface of JFCs cannot be carved on the orbit where these objects are currently observed. 
Finally, following the interplay between surface topography and cometary activity observed on 67P \citep{Vincentetal2019}, \new{\citet{Steckloff2016DPS}, followed by } \citet{Kelley2021} hypothesized that cometary mini-outbursts observed from the ground could be associated with steep terrain features like cliffs and scarps. 

All the aforementioned results converge toward a unique evolutionary sequence evidenced from independent measurables to transform ``young'' cometary surfaces, with sharp surface topography prone to outbursts, into ``old'' cometary surfaces.
Such an evolutionary sequence is of particular interest to the study of Centaurs. Indeed, \citet{guilbert2023pits} suggest that in order to carve the deepest circular pits, water ice should not sublimate otherwise these topographic features would concurrently be eroded. The Centaur phase, experienced during the dynamical evolution of all JFCs, thus appears critical to understand surface properties. The best way to verify the validity of this sequence would be to increase the number of Centaurs and comets for which multiple independent observational techniques are combined (i.e., albedos, phase functions, rotational properties, mini-outburst rates, colors, compositions). For example, a decreasing phase function coefficient would provide a useful observable to characterize the level of erosion of a cometary surface, so would a decreasing rate of observed mini-outbursts. Figure~\ref{fig:surface-prop} indicates that exploring the possible phase-coefficient--albedo correlation for Centaurs is a promising step in this direction. However, more and better observational data are needed to test the evolutionary hypothesis. \new{In Section \ref{sec:outlook} we discuss how the current observing limitations can be overcome in the future using new instruments as well as spacecraft photometric observations covering larger phase angle range inaccessible from the ground.}


\subsubsection{Surface mobilization}
\label{sec:evidence-surface-mobil}

Surface mobilization describes the ensemble of processes that may transport material on an object, locally or globally. The associated physics is typically a competition between the intrinsic properties of the material which tend to keep it in place, and other forces acting on the surface. For instance, landslides occur when the gravitational pull overcomes the material cohesion or the surface is oriented at an angle larger than its \new{critical angle of repose}. In both cases, the material will fail and physically move toward a lower potential energy, downslope.

On active bodies, such as comets and Centaurs, the destabilizing force may come from the sublimation of subsurface volatiles, which may facilitate the surface motion by lowering the material cohesion, or directly accelerate the material against gravity if the gas pressure is significant enough. Other effects such as saltation, electromagnetic forces, or thermal cracking, can also trigger the motion of particles across the surface.

Once mobilized, the material may be redistributed locally, over large distances, or even escape its parent body, depending on the forces at play. This has been well documented on comets, especially with the Rosetta mission which revealed extensive material transport from the southern to the northern regions of the comet, driven by sublimation at perihelion \citep[e.g.,][]{Thomas2015,Davidsson2021b}. Such transport may lead to 
\new{detectable}
color changes (e.g., \citealt{Fornasier2016}) and different surface properties between dust sources, which become depleted and more rock-like on one hand, and the blanketed terrains as depositional sinks on the other \citep{Davidsson2021b}. \new{Due to the lack of 
\final{observational}
evidence 
it is not clear whether similar surface evolution takes place as a result of activity in the Centaur region.}

Although there is no direct evidence for surface mobilization on Centaurs, it is very likely that such process\new{es occur}  since the same forces are at play as on comets or asteroids. In particular, landslides have often been associated with cometary activity and may be triggered by outbursts like those occurring on Centaurs. The literature distinguishes two possible cases: (1) large scale outbursts create deep pits whose vertical side walls are inherently unstable. At any time, additional triggers may lead to further collapse of the walls, concurrently expanding the pit and exposing fresh subsurface material which can then sublimate and lead to more mobilization. This has been invoked as one of the main sources of dust activity on comets \new{\citep{Britt2004, Farnham2013}}, and several cliff collapses were directly observed by Rosetta \citep{Vincent2016, Pajola2017}. 
(2) Even in the absence of cliffs, dusty deposits can still become unstable if their angle of repose becomes too large. This can happen during material deposition, or from a change of local gravitational acceleration that may be triggered by evolution of the rotation state of the body, or tidal forces during a close encounter with larger objects. 
The material thus accelerated may deposit further downslope, or be lofted off by surface activity. This process is likely occurring on comet 103P\final{/Hartley~2} \citep{Steckloff2016}. Laboratory experiments \citep{Kossacki2020, Kossacki2022} have shown that sublimation-triggered landslides can occur on comets on slopes with inclination as low as $10^\circ$. \new{In these experiments, the effect of gas flow within a layer of dust particles is considered. They show that sublimation facilitates sliding and rolling of grains, effectively lowering the critical angle of repose of the material and triggering the sliding slopes that would otherwise be stable. The sliding threshold is discussed through a mobility coefficient, which encompasses multiple physical parameters such as particle size, pore size, dust/ice ratio, temperature, cohesion. If Centaurs are comet-like, with global or local deposits of porous aggregates of dust and ice, this process must be taking place in the giant planet region as well.}


%
%
\section{Outlook to the future}
 
\label{sec:outlook}

As discussed previously in this chapter, our knowledge of Centaurs and, thus, our ability to understand the evolutionary processes at work upon them, is limited by the relative dearth of observations. Owing to their large heliocentric distances, only the largest and/or brightest Centaurs are detected. While most Centaurs are likely rich in volatiles, they reside too far from the Sun for significant water ice sublimation, and many may be depleted in more volatile ices near the surface, limiting or precluding activity. Furthermore, the viewing geometry from Earth changes little, restricting our ability to probe surface properties via changing phase angles. 

However, we are on the cusp of dramatic increases in observational capabilities and survey power, potentially augmented by one or more missions through the outer solar system.  Collectively, these should revolutionize our understanding of Centaurs. We detail below observational advances we expect to most significantly influence our understanding of Centaur evolution over the next decade or two. Undoubtedly, novel data will drive innovation in other areas, such as thermal modeling, dynamical modeling, and laboratory studies. Aside from obvious increases in computational power, the directions that advances in these areas will take are harder to predict and so are not discussed further.





\subsection{Current and very near-future facilities: }

\subsubsection{JWST}

With the successful start of JWST observations in 2022, we are firmly in a new era for solar system studies. At the time of this writing, the first JWST results are just being published (see \fv). JWST is capable of imaging and spectroscopy from 0.6 to 28.5 $\mu$m. Most of this wavelength range is inaccessible from the ground, but is critical for identifying key volatile species expected to be contained in Centaurs, notably H$_2$O, CO, and CO$_2$. JWST's unprecedented sensitivity will allow detection of very low levels of activity in the comae of Centaurs and/or detection of absorption features due to ices on the surface. Detections at a single epoch for a number of Centaurs will reveal the diversity of properties of the population and may yield identifiers that can discriminate past evolutionary histories.  
JWST also has the capability to make coronagraphic images; by occulting a Centaur, deep searches might be conducted for rings or faint binary companions.

The expected lifetime of JWST is 20 years, so Centaurs with semimajor axes of up to 7.4~au could be monitored for an entire orbit. This could be transformative for our understanding of Centaurs' evolution since it would demonstrate how/if the surface properties change with insolation, while a quantification of coma volatile abundances as a function of heliocentric distance would provide unprecedented constraints for numerical models of Centaurs' interiors.

\subsubsection{LSST}
The Vera Rubin Observatory's Legacy Survey of Space and Time (LSST) is scheduled to begin operating in 2025. LSST will survey the southern hemisphere sky every few nights to $r~{\approx}~24.7$ mag, much deeper than existing surveys. Simulations of its detection capabilities \citep{Silsbee2016, Ivezic2019,Schwamb2023} do not specify Centaur discovery rates, but an order of magnitude increase in the number of known Centaurs is consistent with expectations for other small body populations. A debiased analysis of the size-frequency distribution would likely yield significant new insight into how collisionally evolved the population is and to what extent the population is modified as objects transition from TNOs to JFCs.
LSST will obtain data in \new{the standard broadband} {\it u, g, r, i, z}, and {\it y} filters,
allowing robust spectro-photometric color measurements to be made of the Centaur population. When compared with other physical properties and dynamical simulations, these colors might yield insight into past dynamical histories and links with other well-characterized populations. 

Depending on the brightness around the orbit \new{and the extent to which LSST observes the northern part of the ecliptic \citep[the survey design is not yet finalized; see discussion in][]{Schwamb2023}}, LSST will \new{likely} detect each Centaur tens to a few hundred times during its planned 10-year operating time, which should allow rotational lightcurves to be measured or constrained. Given how few Centaur rotation periods and axial ratio measurements currently exist, even poorly constrained lightcurves will be informative for understanding the prevalence of highly elongated objects \final{\citep{Donaldson2023,Donaldson2024}}
and associated implications for Centaurs' internal strength. The regular observations will also yield robust limits on activity and the frequency of outbursts. While 29P is well-known for its frequent outbursts and sustained activity \citep[e.g.,][]{Miles2016,Wierzchos2020}, it is currently unknown how common such behavior is across the Centaur population. As with other properties just discussed, comparison of activity and outburst rates with other physical and dynamical properties may yield insight into the mechanisms at work in Centaurs.

\subsubsection{\new{SPHEREx}}

\new{NASA's SPHEREx \citep[Spectro-Photometer for the History for the Universe, Epoch of Reionization, and Ices Explorer][]{Dore2018}, is a space-based all-sky spectral survey from 0.75 to 5.0 $\mu$m due to launch in 2025. Its core mission will survey the entire sky every six months for two years, collecting data in 96 spectral channels for each pixel to approximately the same limiting magnitude as WISE \citep{Ivezic2022}. Owing to the instrument design, the spectral channels will not be obtained simultaneously, but over several days for a given object. Thus, it will likely provide low resolution ($R=35-130$), rotationally smeared spectra for several dozen Centaurs (based on WISE discovery rates; \citealt{Bauer2013}). Since this wavelength region is diagnostic of the presence of various key ices (see Section~\ref{sec:surface_properties}), SPHEREx should be a valuable complement to targeted observations by JWST (notably the DiSCo-TNOs large program) for investigating compositional differences between the TNO, Centaur, and comet populations. 
SPHEREx contains no major expendables \citep{Lisse2024}, so it could observe significantly longer than the core mission, potentially allowing shift-and-stack detection of fainter objects, or catching more objects as they brighten near perihelion.}


\subsection{Future telescopes}

Looking beyond JWST, LSST, \new{and SPHEREx} several next-generation facilities have sufficiently developed plans to warrant discussion. First, up to three 30-m class ground-based optical telescopes are likely to become operational in the next decade.
With mirror diameters roughly three times as large as today's largest facilities, these will have unrivaled sensitivity and angular resolutions. Most compellingly, the diffraction-limited angular resolutions of these facilities will be $0.005-0.010$~arcsec, which will result in \new{some} of the larger\new{/closer} Centaurs subtending more than one pixel on the detector. \new{Deconvolution techniques \citep[e.g.,][]{Marchis2006} should allow at least hints of the shapes for Centaurs like 29P (diameter of $\sim$0.014~arcsec at perihelion) and Chiron (diameter of $\sim$0.027~arcsec at perihelion), and might allow the identification of close binaries \citep[cf.][]{Noll2008,Agarwal2017}.} 
Thus, it will be possible to measure some Centaur nucleus shapes directly\new{,} to make rotationally resolved maps of their surface properties\new{, and to quantify the frequency of close binaries}.

Second, NASA's planned NEO Surveyor will be a space-based IR (4 to 10 $\mu$m) survey telescope designed to detect the thermal emission from near-Earth objects. Designed as a more powerful successor to WISE/NEOWISE (25\% larger mirror), NEO Surveyor will be sensitive to a narrower spectral range that the WISE primary mission, but a wider spectral range than NEOWISE, and is expected to operate  for 12 years as compared to WISE's 10 months. WISE constrained sizes and albedos for 41 Centaurs \citep{Bauer2013}; NEO Surveyor should obtain comparable data for smaller, darker, and/or more distant Centaurs. This will allow the size-frequency distribution to be extended closer to the sizes of comet nuclei, thus probing the extent to which the Centaur population has been collisionally modified and thereby constraining models of solar system evolution \citep[e.g.,][]{Bottke2023}. Comparison of the albedo with other photometric properties such as color and phase function will yield new insight into the processes affecting Centaur surfaces.

Finally, several space-based facilities that might contribute new knowledge about Centaurs have either recently launched (ESA's Euclid) or are planned to launch by the end of the decade (NASA's Nancy Grace Roman, ESA's PLATO, and China's Xuntian). All will operate at optical or IR wavelengths, where Centaurs can be best studied. However, since none will prioritize solar system science, their major contributions to our understanding of Centaurs is likely to be via serendipitous observations \citep[cf.][]{Carry2018}. These are likely to include new discoveries, spectrophotometric measurements, and possibly detections of wide binaries. Roman may support a guest investigator program; ideas for Centaur science are discussed in \citet{Holler2018} and include detection of faint activity, measurement of colors, and IFU spectroscopy. Roman will have a coronagraphic imager, potentially enabling searches for rings or faint companions.




\subsection{\new{Ongoing and future space missions} }

Launched in 2021, NASA's Lucy mission is on a 12-year journey to visit both of Jupiter's Trojan asteroid populations \citep{Levison2021}. Since the Centaur region is likely populated with some asteroids that leaked from Jupiter Trojan orbits (see Section~\ref{sec:life-cycle}), high resolution observations of the seven Trojans that Lucy will visit should provide robust information about this population. This may prove useful for interpreting and contextualizing Centaur observations, e.g., potentially identifying observable properties by which escaped Trojans can be identified within the Centaur population. Furthermore, observations of Centaurs during the nominal mission or during a potential extended mission would be capable of sampling Centaurs at larger phase angles than is possible from Earth. Observations acquired beyond the opposition effect region will be sensitive to the surface topography, as the effects of shadows on the photometry become more pronounced. Such observations would allow comparison with comet nuclei 
and the TNOs that have been sampled in a similar manner by the New Horizons mission \citep{Porter2016,Verbiscer2019,Verbiscer2022}, further illuminating our understanding of how surface roughness evolves.

ESA's Comet Interceptor mission is slated to launch around 2029, and will fly-by a yet to be discovered dynamically new or returning Oort cloud comet in the early 2030s \final{\citep{Jones2019,Jones2024}}. All previous missions to comets have visited Jupiter-family or Halley-type comets which have been through the inner solar system many times. Comparison of the surface properties of the Comet Interceptor target with past mission targets will give novel insight into how comet surfaces evolve. As Centaurs are at an intermediate evolutionary step, these results will provide context for interpreting Centaur surface properties, and may yield insight into how to use observable surface properties as diagnostics of Centaurs' evolutionary differences, e.g., identifying ones that have previously been at substantially smaller heliocentric distances.




\subsection{Potential future space missions}

Several space missions that could influence our understanding of Centaur evolution might plausibly be selected in the next decade. The most obviously relevant mission would be a mission to a Centaur. Two such missions were proposed in the 2019 NASA Discovery mission call\new{, Centaurus \citep{Singer2019} and Chimera \citep{Harris2019}}, and the NRC Planetary Science Decadal Survey 2023-2032 
ranked a Centaur orbiter and lander among the top seven mission themes for the New Frontiers 6 call. See \cite{Harris2025} for further discussion \new{on the motivation and science goals of the proposed mission concepts}.

A Uranus orbiter and probe, to launch in the 2030s, was the highest priority flagship mission by the 2023-2032 Decadal survey \citep{Mandt2023}, while ESA's Voyage 2050 recommended a mission to the moons of giant planets as a top three priority for a future large-class mission. An advanced mission concept for an interstellar probe has been developed for the next U.S.\ Solar and Space Physics Decadal Survey, and ESA's Voyage 2050 advocated for potential participation.
This mission would launch in 2036--2041 and rapidly leave the solar system, traveling at 6--7 au per year. Any of these missions would necessarily traverse the Centaur region and could potentially be designed to fly-by one or more Centaurs en route. Even without a fly-by encounter, remote observations of Centaurs during the cruise phase of a mission would sample an even larger range of phase angles than could the Lucy mission (see previous subsection), and might be more sensitive to gas and dust activity than Earth-based observations.

%
%
\section{Concluding remarks}
\label{sec:conclusion}

After the discovery of the Kuiper Belt and the recognition of the TNO-Centaur-JFC continuum, the Centaur population was largely regarded as the transition phase experienced by TNOs on their way \new{to} becoming JFCs. Growing interest in the unique properties of Centaurs, however, has led to more thorough studies, which have unveiled them as highly interesting and complex in their own right. First, the dynamical evolution of Centaurs into JFCs is now seen as an intricate dance combining alternating residences in the colder Centaur region with periods as JFCs. Moreover, even though most Centaurs are sourced from the Scattered Disk, other populations throughout the solar system have been identified as potential contributors to the Centaur region. Finally, the time spent on orbits in the giant planet region has been found to lead to significant evolution of these objects. We now \new{think} that characterizing the evolution in the Centaur region is crucial for understanding the life cycle of outer-solar-system planetesimals and for drawing accurate conclusions about the early stages of the solar system. 

The main goal of this chapter has been to review the current understanding of the processes driving the current-epoch Centaur evolution in the giant planet region. We set out to describe the mechanisms shaping Centaur evolution and the evidence that informs our understanding of these processes. At a quick glance, the rapidly growing body of work we have reviewed paints a relatively straightforward picture: the driving force behind the changes experienced by Centaurs is orbital evolution. Their orbital dynamics is what controls the environmental factors that enable thermal, collisional\new{,} or tidal-force processing to take place. However, a closer look at individual objects instantly reveals a much greater complexity. The extent to which an object gets altered depends on the intricate interplay of its physical properties, structure and composition. As one of the most diverse populations in the solar system, Centaurs thus pose significant challenges 
\new{to identifying the}
unique signatures of evolution. Nevertheless, we are optimistic that 
\new{these obstacles}
will be 
\new{overcome}
with the spectacular observing prospects presented in Section~\ref{sec:outlook}. These expected advances make us confident that the coming decades will offer remarkable progress in our understanding of Centaurs and their related populations, which will enable us to acquire a thorough insight into their evolution in the giant planet region.


\vskip .2in
\noindent \textbf{Acknowledgments} 

This work was was supported by the International Space Science Institute (ISSI) in Bern, through ISSI International Team project 504 ``The Life Cycle of Comets". RK would like to acknowledge the support from ``L’Oreal UNESCO For Women in Science" National program for Bulgaria. AGL received funding from the European Research Council (ERC) under the European Union’s Horizon 2020 research and innovation programme (Grant Agreement No 802699).


\bibliographystyle{plainnat}

\bibliography{References}

\end{document}